\input harvmac
\input amssym.def
\input amssym.tex
\noblackbox

\newif\iflanl
\openin 1 lanlmac
\ifeof 1 \lanlfalse \else \lanltrue \fi
\closein 1
%
\newif\ifhypertex
\ifx\hyperdef\UnDeFiNeD
    \hypertexfalse
    \message{[HYPERTEX MODE OFF}
    \def\hyperref#1#2#3#4{#4}
    \def\hyperdef#1#2#3#4{#4}
    \def\hypernoname{}
    \def\e@tf@ur#1{}
    \def\hth/#1#2#3#4#5#6#7{{\tt hep-th/#1#2#3#4#5#6#7}}
    
\else
    \hypertextrue
    \message{[HYPERTEX MODE ON}
  \def\hth/#1#2#3#4#5#6#7{
  {\tt hep-th/#1#2#3#4#5#6#7}}

\fi
\newif\ifdraft

\noblackbox
\catcode`\@=11
\newif\iffrontpage
%
%
%
%
\newif\iffigureexists
\newif\ifepsfloaded
\def\epsfcheck{
\ifdraft
\input epsf\epsfloadedtrue
\else
  \openin 1 epsf
  \ifeof 1 \epsfloadedfalse \else \epsfloadedtrue \fi
  \closein 1
  \ifepsfloaded
    \input epsf
  \else
\immediate\write20{NO EPSF FILE --- FIGURES WILL BE IGNORED}
  \fi
\fi
\def\epsfcheck{}}
\def\checkex#1{
\ifdraft
\figureexistsfalse\immediate%
\write20{Draftmode: figure #1 not included}
\figureexiststrue 
\else\relax
    \ifepsfloaded \openin 1 #1
        \ifeof 1
           \figureexistsfalse
  \immediate\write20{FIGURE FILE #1 NOT FOUND}
        \else \figureexiststrue
        \fi \closein 1
    \else \figureexistsfalse
    \fi
\fi}
\def\missbox#1#2{$\vcenter{\hrule
\hbox{\vrule height#1\kern1.truein
\raise.5truein\hbox{#2} \kern1.truein \vrule} \hrule}$}
\def\lfig#1{
\let\labelflag=#1%
\def\numb@rone{#1}%
\ifx\labelflag\UnDeFiNeD%
{\xdef#1{\the\figno}%
\writedef{#1\leftbracket{\the\figno}}%
\global\advance\figno by1%
}\fi{\hyperref{}{figure}{{\numb@rone}}{Fig.{\numb@rone}}}}
\def\figinsert#1#2#3#4{
\epsfcheck\checkex{#4}%
\def\figsize{#3}%
\let\flag=#1\ifx\flag\UnDeFiNeD
{\xdef#1{\the\figno}%
\writedef{#1\leftbracket{\the\figno}}%
\global\advance\figno by1%
}\fi
\goodbreak\midinsert%
\iffigureexists
\centerline{\epsfysize\figsize\epsfbox{#4}}%
\else%
\vskip.05truein
  \ifepsfloaded
  \ifdraft
  \centerline{\missbox\figsize{Draftmode: #4 not included}}%
  \else
  \centerline{\missbox\figsize{#4 not found}}
  \fi
  \else
  \centerline{\missbox\figsize{epsf.tex not found}}
  \fi
\vskip.05truein
\fi%
{\smallskip%
\leftskip 4pc \rightskip 4pc%
\noindent\ninepoint\sl \baselineskip=11pt%
{\bf{\hyperdef\hypernoname{figure}{{#1}}{}}~}#2%
\smallskip}\bigskip\endinsert%
}

\newif\ifdraft

\catcode`\@=11
\newif\iffrontpage

\noblackbox

\def\frac#1#2{{#1\over #2}}

\def\{{\lbrace}
\def\}{\rbrace}

\def\C{{\Bbb C}}

\def\P{{\Bbb P}}
\def\F{{\Bbb F}}

\def\a{\alpha}

\def\l{\lambda}

\def\p{\partial}
%

%
\def\abstract#1{
\vskip.5in\vfil\centerline
{\bf Abstract}\penalty1000
{{\smallskip\ifx\answ\bigans\leftskip 2pc \rightskip 2pc
\else\leftskip 5pc \rightskip 5pc\fi
\noindent\abstractfont \baselineskip=12pt
{#1} \smallskip}}
\penalty-1000}
\def\tr{{\rm tr}}

\vbadness=10000

%
\lref\AKMV{M.~Aganagic, A.~Klemm, M.~Mari\~no and C.~Vafa,
``Matrix model as a mirror of Chern-Simons theory,''
hep-th/0211098.}
%
\lref\Akemann{
G.~Akemann,
``Higher genus correlators for the Hermitian matrix model with multiple cuts,''
Nucl.\ Phys.\ B {\bf 482}, 403 (1996),
hep-th/9606004.
}
%
%
\lref\AMV{
M.~Aganagic, M.~Mari\~no and C.~Vafa,
``All loop topological string amplitudes from Chern-Simons theory,''
hep-th/0206164.
}
%

\lref\ACKM{
J.~Ambjorn, L.~Chekhov, C.~F.~Kristjansen and Y.~Makeenko,
``Matrix model calculations beyond the spherical limit,''
Nucl.\ Phys.\ B {\bf 404}, 127 (1993),
Erratum-ibid.\ B {\bf 449}, 681 (1995),
hep-th/9302014.
}
%
\lref\CM{
L.~Chekhov and A.~Mironov,
``Matrix models vs. Seiberg-Witten/Whitham theories,''
hep-th/0209085.
}
%
\lref\BCOVone{
M.~Bershadsky, S.~Cecotti, H.~Ooguri and C.~Vafa,
``Holomorphic anomalies in topological field theories,''
Nucl.\ Phys.\ B {\bf 405}, 279 (1993)
hep-th/9302103.
}
%
\lref\BCOVtwo{
M.~Bershadsky, S.~Cecotti, H.~Ooguri and C.~Vafa,
``Kodaira-Spencer theory of gravity and exact results for
quantum string amplitudes,''
Commun.\ Math.\ Phys.\  {\bf 165}, 311 (1994),
hep-th/9309140.
}
\lref\BF{ P.F.~Byrd and M.D.~Friedman, {\sl Handbook of Elliptic
Integrals for Engineers and Scientists}, Springer Verlag, New York
1971.}
%
\lref\BDE{ G.~Bonnet, F.~David and B.~Eynard, ``Breakdown of
universality in multi-cut matrix models,'' J.\ Phys.\ A {\bf 33},
6739 (2000), cond-mat/0003324.
}
%
\lref\CKYZ{
T.~M.~Chiang, A.~Klemm, S.~T.~Yau and E.~Zaslow,
`Local mirror symmetry: Calculations and interpretations,''
Adv.\ Theor.\ Math.\ Phys.\  {\bf 3}, 495 (1999),
hep-th/9903053.
}
%
\lref\DVone{
R.~Dijkgraaf and C.~Vafa,
``Matrix models, topological strings, and supersymmetric gauge theories,''
Nucl.\ Phys.\ B {\bf 644}, 3 (2002),
hep-th/0206255.
}
%
\lref\DVtwo{
R.~Dijkgraaf and C.~Vafa,
``On geometry and matrix models,''
Nucl.\ Phys.\ B {\bf 644}, 21 (2002),
hep-th/0207106.
}
%
\lref\DVthree{
R.~Dijkgraaf and C.~Vafa,
``A perturbative window into non-perturbative physics,''
hep-th/0208048.
}
%
\lref\DGKV{
R.~Dijkgraaf, S.~Gukov, V.~A.~Kazakov and C.~Vafa,
``Perturbative analysis of gauged matrix models,''
hep-th/0210238.
}
%
\lref\CIV{ F.~Cachazo, K.~A.~Intriligator and C.~Vafa, ``A large
$N$ duality via a geometric transition,'' Nucl.\ Phys.\ B {\bf
603}, 3 (2001), hep-th/0103067.
}
%
\lref\CV{
F.~Cachazo and C.~Vafa,
``${\cal N}=1$ and ${\cal N}=2$ geometry from fluxes,''
hep-th/0206017.
}
%
\lref\Flume{
R.~Flume and R.~Poghossian,
``An algorithm for the microscopic evaluation of the coefficients
of the  Seiberg-Witten prepotential,''
hep-th/0208176.
}
%
\lref\Hosono{
S.~Hosono,
``Counting BPS states via holomorphic anomaly equations,''
hep-th/0206206.
}
\lref\KKLM{S.~Kachru, S.~Katz, A.~E.~Lawrence and J.~McGreevy,
``Open string instantons and superpotentials,'' Phys.\ Rev.\ D
{\bf 62}, 026001 (2000), hep-th/9912151.}

\lref\KKLMV{ S.~Kachru, A.~Klemm, W.~Lerche, P.~Mayr and C.~Vafa,
``Nonperturbative results on the point particle limit of ${\cal
N}=2$ heterotic string compactifications,'' Nucl.\ Phys.\ B {\bf
459}, 537 (1996), hep-th/9508155.
}
\lref\KKV{
S.~Katz, A.~Klemm and C.~Vafa,
``Geometric engineering of quantum field theories,''
Nucl.\ Phys.\ B {\bf 497}, 173 (1997);
hep-th/9609239.
}
%
%
\lref\KKVtwo{ S.~Katz, A.~Klemm and C.~Vafa, ``M-theory,
topological strings and spinning black holes,'' Adv.\ Theor.\
Math.\ Phys.\  {\bf 3}, 1445 (1999), hep-th/9910181.
}
%
\lref\KMV{
S.~Katz, P.~Mayr and C.~Vafa,
``Mirror symmetry and exact solution of 4D N = 2 gauge theories. I,''
Adv.\ Theor.\ Math.\ Phys.\  {\bf 1}, 53 (1998)
[arXiv:hep-th/9706110].
}
\lref\KZ{
A.~Klemm and E.~Zaslow,
``Local mirror symmetry at higher genus,''
hep-th/9906046.
}
%
\lref\KLT{ A.~Klemm, W.~Lerche and S.~Theisen, ``Nonperturbative
effective actions of ${\cal N}=2$ supersymmetric gauge theories,''
Int.\ J.\ Mod.\ Phys.\ A {\bf 11}, 1929 (1996), hep-th/9505150.
}
\lref\NSW{S.~Naculich, H.~Schnitzer and N.~Wyllard,
``The ${\cal N} = 2$ $U(N)$ gauge theory prepotential and
periods from a  perturbative matrix model calculation,'' hep-th/0211123.}

\lref\OV{ H.~Ooguri and C.~Vafa, ``Worldsheet derivation of a
large $N$ duality,'' Nucl.\ Phys.\ B {\bf 641}, 3 (2002),
hep-th/0205297.
}
%
\lref\MW{ G.~Moore and E.~Witten, ``Integration over the $u$-plane
in Donaldson theory,'' Adv.\ Theor.\ Math.\ Phys.\  {\bf 1}, 298
(1998), hep-th/9709193.
}
\lref\others{
A.~Losev, N.~Nekrasov and S.~L.~Shatashvili,
``Issues in topological gauge theory,''
Nucl.\ Phys.\ B {\bf 534}, 549 (1998),
hep-th/9711108. M.~Mari\~no and G.~Moore,
``The Donaldson-Witten function for gauge groups of rank larger than one,''
Commun.\ Math.\ Phys.\  {\bf 199}, 25 (1998),
hep-th/9802185.
}
%
\lref\Nekrasov{
N.~A.~Nekrasov,
``Seiberg-Witten prepotential from instanton counting,''
hep-th/0206161.
}
%
\lref\SW{
N.~Seiberg and E.~Witten,
``Electric - magnetic duality, monopole condensation, and confinement in
${\cal N}=2$ supersymmetric Yang-Mills theory,''
Nucl.\ Phys.\ B {\bf 426}, 19 (1994),
Erratum-ibid.\ B {\bf 430}, 485 (1994),
hep-th/9407087.
}
\lref\SWtwo{N.~Seiberg and E.~Witten,
``Monopoles, duality and chiral symmetry breaking in ${\cal N}=2$
supersymmetric QCD,''
Nucl.\ Phys.\ B {\bf 431}, 484 (1994), hep-th/9408099.}
\lref\Witten{ E.~Witten, ``On $S$ duality in Abelian gauge
theory,'' hep-th/9505186.
}
%
\lref\VW{ C.~Vafa and E.~Witten, ``A strong coupling test of $S$
duality,'' Nucl.\ Phys.\ B {\bf 431}, 3 (1994), hep-th/9408074.
}
%
\lref\Vafa{ C.~Vafa,
``A stringy test of the fate of the
conifold,'' Nucl.\ Phys.\ B {\bf 447}, 252 (1995), hep-th/9505023.
}

\lref\VafaT{ C.~Vafa, ``Superstrings and topological strings at
large $N$,'' J.\ Math.\ Phys.\  {\bf 42}, 2798 (2001),
hep-th/0008142.
}

\Title{\vbox{ \rightline{\vbox{\baselineskip12pt
\hbox{AEI-2002-090} \hbox{HUTP-02/A060} \hbox{HU-EP-02/48}
\hbox{hep-th/yymmddd}}}}} {\vbox{\centerline{Gravitational
corrections in supersymmetric gauge theory}
\smallskip
\centerline{and matrix models}}}
\vskip0.3cm
\centerline{ Albrecht~Klemm$^a$, Marcos~Mari\~no$^b$ and
Stefan~Theisen$^c$ }
\vskip 0.6cm
\centerline{$^a$ \it Humboldt
Universit\"at, Institut f\"ur Physik, Berlin, Germany}
\vskip.2cm
\centerline{$^b$\it Jefferson Physical Laboratory, Harvard
University, Cambridge, MA 02139, USA }
\vskip.2cm
\centerline{$^c$
\it Max-Planck-Institut f\"ur Gravitationsphysik,
Albert-Einstein-Institut, Golm, Germany}
\vskip 0.0cm
\abstract{
Gravitational corrections in ${\cal N}=1$ and ${\cal N}=2$
supersymmetric gauge 
theories are obtained from topological string amplitudes.  
We show how they are recovered in matrix model computations. 
This provides a test of the proposal by Dijkgraaf and Vafa
beyond the planar limit. 
Both, matrix model and topological string theory, are used 
to check a conjecture of Nekrasov concerning these
gravitational couplings in Seiberg-Witten theory. 
Our analysis is performed for those gauge theories which are 
related to the cubic matrix model, i.e. pure $SU(2)$ 
Seiberg-Witten theory and ${\cal N}=2$ $U(N)$ SYM broken to 
${\cal N}=1$ via a cubic superpotential. 
We outline the computation of the topological amplitudes 
for the local Calabi-Yau manifolds which are relevant for 
these two cases. 
}
\Date{\vbox{\hbox{\sl {November 2002}} }}
\goodbreak

\parskip=4pt plus 15pt minus 1pt
\baselineskip=15pt plus 2pt minus 1pt

\newsec{Introduction}

The fact that string theory can be a powerful tool to study
four-dimensional supersymmetric gauge theories has been long
appreciated. One of the most successful approaches is geometric
engineering, where the gauge theory is appropriately embedded
in a local Calabi-Yau compactification of type II string theory. 
A large and interesting class of supersymmetric field theories can be
geometrically engineered in this way, including ${\cal N}=2$
supersymmetric Yang-Mills theory with or without matter~\KKV\KMV . 
The gauge theory is recovered in a certain singular limit of 
the Calabi-Yau geometry where the string corrections decouple, 
and the gauge group and matter content of the field theory depend 
only on local properties of the Calabi-Yau space near the singularity.

Type II string theory on Calabi-Yau geometries has a topological
subsector called topological string theory. It turns out that
amplitudes in topological string theory compute exactly certain
holomorphic terms in the effective action \BCOVtwo. There are two
types of topological strings, the A-model and the B-model. In the
A-model, the relevant correlation functions are sums of
world-sheet instanton contributions of a given genus. Using mirror
symmetry one can relate the A-model to the B-model, and the latter
provides an elegant and efficient tool to perform these sums
exactly. From a pure field theory point of view the exact
holomorphic information comes from summing over space-time
instantons.
The decoupling limit relates string world-sheet instantons to
field theory space-time instantons, and for example the space-time
instanton corrected exact gauge coupling of ${\cal N}=2$
Seiberg-Witten theory \refs{\SW,\SWtwo} is recovered in that limit
from the genus zero topological string amplitude. Higher genus
world-sheet instanton corrections are believed to calculate
exactly, in the same limit, certain (holomorphic) couplings of the
gauge theory to gravity. On the field theory side, these couplings
haven't been much studied. For ${\cal N}=2$ theories, the genus
one coupling can be extracted from the low-energy twisted theory,
where the coupling to gravity plays an important role
\refs{\Witten,\VW,\MW,\others}. More recently there has been
important progress in calculating also the space-time instantons
contributions for higher genus \refs{\Nekrasov,\Flume}.

In the context of geometric engineering of gauge theories, one can
break ${\cal N}=2$ to ${\cal N}=1$ by adding space-time filling
D-branes wrapped on internal supersymmetric cycles, or in a dual
picture by turning on RR-fluxes. The best studied holomorphic
${\cal N}=1$ quantity is the superpotential, which is computed via
genus zero open or closed topological string amplitudes,
respectively. As in the ${\cal N}=2$ case, higher genus amplitudes
give gravitational corrections to the gauge theory effective
action \refs{\BCOVtwo,\VafaT}.

Recently Dijkgraaf and Vafa \refs{\DVone,\DVtwo,\DVthree} have
made the exciting proposal that these holomorphic terms in ${\cal
N}=1$ and ${\cal N}=2$ supersymmetric field theories are
calculated by matrix models. The genus expansion is replaced by an
expansion in $1/N$, $N$ being the dimension of the matrices. This leads to
the possibility to use matrix models techniques to study
supersymmetric gauge theories as well as topological string
theories. In particular in the ${\cal N}=1$ context the tree level
superpotential appears in the action of the matrix model and the
exact effective low-energy superpotential is recovered in the
planar limit. This has been checked explicitly in various
examples, see for example \refs{\DVthree,\DGKV}. Although the
original proposal of Dijkgraaf and Vafa applies to ${\cal N}=1$
models, one can recover results for the ${\cal N}=2$ theory in an
appropriate limit \refs{\CV,\DVthree,\DGKV}.

The aim of this paper is to bring the above ideas together in
order to perform a non-trivial check of the matrix model proposal
beyond the planar limit, calculating thereby the corresponding
gravitational terms in the supersymmetric field theories with
completely independent methods. In this paper we will consider the
simplest case of the cubic matrix potential. According to
Dijkgraaf and Vafa, this matrix model captures the holomorphic couplings of
IIB theory on a Calabi-Yau
geometry with fluxes that generate a superpotential. On the other
hand, ${\cal N}=2$ $SU(2)$ Seiberg-Witten theory is recovered at a
special co-dimension one sub-locus in the moduli space, where the
${\cal N}=2$ supersymmetry is restored. This in turn can be
geometrically engineered by decoupling the string correction from
the dual type IIA string on the canonical line bundle over
$\P^1\times \P^1=\F_0$ \KKV. The interrelations between these theories
are shown in the diagram.

\figinsert\figone{}{2.4in}{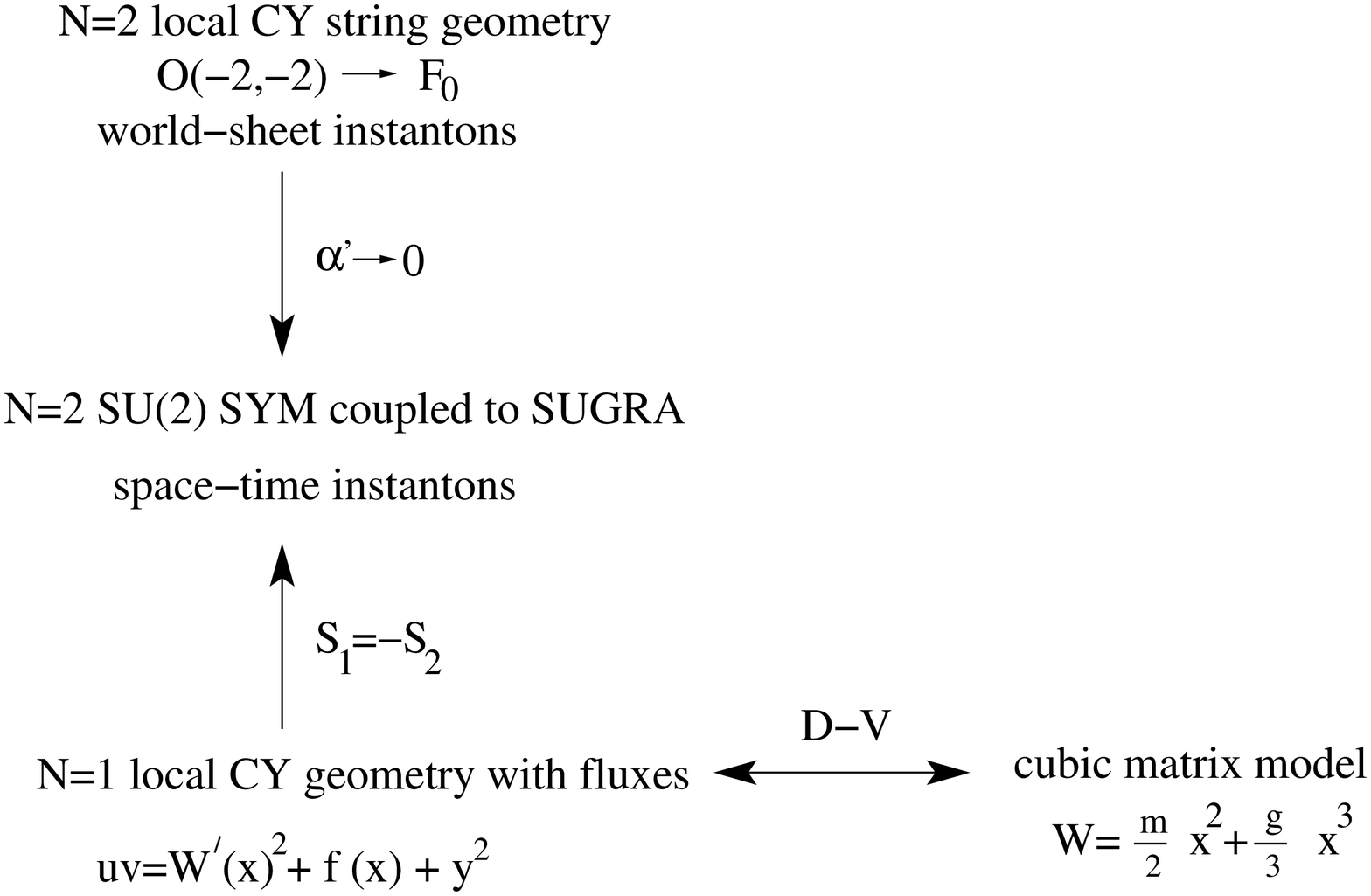}

We are able to check non-planar contributions of the cubic matrix
model against the holomorphic gravitational couplings in various
ways. First, we calculate the exact genus one free-energy
$F^{(1)}$ of the matrix model over the whole ${\cal N}=1$ moduli
space. This matches the genus one contribution in the
corresponding IIB model, which we obtained using the holomorphic
anomaly equation. Second, we consider the holomorphic
gravitational couplings of the ${\cal N}=2$ theory. We compute
them in the string decoupling limit of type IIA theory on local
$F_0$, and we check these in this limit against the corresponding
space-time instanton calculations. Finally, we show that, by
specializing the matrix model answer for $F^{(1)}$ to the ${\cal
N}=2$ subspace, one recovers the right answer.

The paper is organized as follows. In section 2, we analyze
topological string theory on the relevant local Calabi-Yau
geometries: local $\F_0$, which geometrically engineers ${\cal
N}=2$, $SU(2)$ Yang-Mills theory, and the geometries that engineer
${\cal N}=2$ $U(N)$ super Yang-Mills theory broken down to ${\cal
N}=1$ by a tree level cubic superpotential $W(\Phi)$. In section
3, we derive an expression for the gravitational correction to
$F^{(1)}$ of ${\cal N}=2$, $SU(2)$ Yang-Mills theory from
topological field theory, and we show that it agrees with the
decoupling limit of the corresponding amplitude on local $\F_0$.
The geometries considered in section 2 are both captured by the
cubic matrix model, which we analyze in section 4 from a
perturbative point of view and also from the point of view of the
loop equations. In particular, we derive an exact expression for
$F^{(1)}$ that is shown to agree with the one computed in section
2 for the ${\cal N}=1$ theory. We also analyze the embedding of
${\cal N}=2$ in the matrix model and we reproduce the
gravitational coupling of $SU(2)$ Seiberg-Witten theory from the
genus one free energy of the matrix model. In section 5 we focus again on
$SU(2)$, ${\cal N}=2$ and we show that the results obtained in
this paper agree with the direct instanton computations of
\refs{\Nekrasov, \Flume}, confirming in this way a conjecture of
Nekrasov.

\newsec{Topological string in the associated local Calabi-Yau
geometries}

In this section we will consider topological string theory on two
different pairs of non-compact Calabi-Yau three-folds $I$ and
$II$. They are both related to the cubic matrix model in a way to
be described below.

The first pair of Calabi-Yau three-folds, denoted by $I$, is a
mirror pair. The A-model geometry of $I$ is the total space of the
anti-canonical line bundle ${\cal O}(-2,-2)\rightarrow \P^1\times
\P^1$. The B-model is its local mirror geometry, which is (in a
patch) given by the constraint
\eqn\mirrorPIxPI{vw=1+Y_1+{e^{-\hat t_1}\over Y_1}
+Y_2+{e^{-\hat t_2}\over Y_2}\, }
where $v,w\in \C$ and $Y_1,Y_2\in \C^*$. $\hat t_i$ are the two
complex structure deformations. A canonical parameterization of
the complex structure moduli space is in terms of periods $t_i$ of
a meromorphic differential on the elliptic curve obtained from
\mirrorPIxPI\ by setting $vw=0$~\KKV. The relation between $t_i$
and $\hat t_i$ will be given below. The Picard-Fuchs equations as
well as explicit expressions for the periods in terms of various
expansion parameters can be found in \refs{\KKV,\AKMV}.

The Calabi-Yau three-folds in the second pair, which will be
denoted by $II$ and $\widehat {II}$, are related by a geometric
transition. The $II$ geometry is obtained by deforming the bundle
${\cal O}(-2)\oplus {\cal O}(0) \rightarrow \P^1$, such that the
location of the $\P^1$ section in the ${\cal O}(0)$ direction,
parameterized by $x$, is restricted to the two critical points,
$a_1,a_2$, of a cubic potential $W(x)={m\over 2} x^2 + {g\over 3}
x^3$ \CIV. Consider now type IIB string theory on this Calabi-Yau
three-fold. As pointed out in \refs{\KKLM,\CIV}, by wrapping
$N_1$, $N_2$ D5 branes around the $\P_1$'s located at $a_1$,
$a_2$, respectively, one can geometrically engineer an ${\cal
N}=1$ theory. This theory is $U(N)$ ${\cal N}=2$ Yang-Mills theory
broken to ${\cal N}=1$ via the addition of the superpotential
$W(\Phi)$ for the ${\cal N}=1$ adjoint chiral superfield $\Phi$
that is part of the ${\cal N}=2$ gauge multiplet. The
configuration with the above distribution of branes corresponds to
a classical vacuum where $U(N)\rightarrow U(N_1)\times U(N_2)$. At
low energies the $SU(N_i) \subset U(N_i)$ parts confine and the
unbroken gauge group is a product of $U(1)$ factors. For each
$SU(N_i)$ one has, at low energy, a glue-ball superfield $S_i$.
The resulting dynamics is governed by the effective superpotential
for the glue-balls, $W_{\rm eff}(S_i)$.

The $\widehat {II}$ geometry is obtained from $II$ by a geometric
transition \CIV, in which the two $\P^1$'s are contracted to a
point and the resulting singular space is smoothened by two
$S^3$'s of finite size. This has the local description as a
hypersurface in $\C^4$:
\eqn\transciv{vw=W'(x)^2+f_1(x)+y^2,}
where $x,y,v,w\in \C$ and $f_1$ is a polynomial of order one,
which splits the double zeros of $W'(x)^2$ to $a_1^\pm,a_2^\pm$.
The two complex structure deformations of \transciv\ are the two
parameters of $f_1$ or, alternatively, the differences
$a_i^+-a_i^-$. The periods $S_i={1\over 2 \pi
i}\int_{a_i^-}^{a_i^+} \omega $ and $\Pi_{i}={1\over 2 \pi i}
\int_{a_i^+}^{\Lambda} \omega$, $i=1,2$, where $\omega={\rm
d}x\sqrt{W'(x)^2+f_1(x)}$, emerge by integrating over two
dimensions of the period integrals of the holomorphic three-form
of the local Calabi-Yau geometry \transciv\CIV. They are functions
of the complex structure parameters and of the parameters $m$ and
$g$ which appear in $W$. The $\Pi_i$ also depend on a cut-off
$\Lambda$ which must be introduced since we are working on a
non-compact curve (i.e. $x\in\C$ rather than $x\in\P^1$; otherwise
there would be e.g. a linear relation $S_1+S_2=0$). After the
geometric transition the D5 branes disappear, and we are left with
a closed IIB string geometry with fluxes.

\subsec{Topological amplitudes on ${\cal O}(-2,-2)\rightarrow \P^1\times \P^1$
and their field theory interpretation.}

The geometry  $I$ was considered in \KKV\ to geometrically
engineer $SU(2)$ Seiberg-Witten field theory from string theory.
The two moduli of the A-model are the complexified K\"ahler
parameters $t_1$ and $t_2$ of the two $\P^1$'s. String corrections
to the gauge theory can be decoupled by sending
$\epsilon\sim\sqrt{\a'}\to 0$. This should be done in such a way
as to preserve the renormalization group relation $t_1\sim 1/g_s^2
\sim 1/g_{\rm YM}^2\sim\log({m_W\over\Lambda})$ where $m_W\sim
a\sim t_2$, $\Lambda$ is the scale of the gauge theory, $g_s$ the
string coupling and $a$ parameterizes the expectation value of the
adjoint scalar field in the Cartan subalgebra of $SU(2)$. This
leads to the double scaling limit \foot{In the heterotic--type II
duality $t_1$ is identified with the size of the base of the K3
fibration and $t_2$ with the size of a 2-cycle in the fiber.}
$\epsilon\rightarrow 0$ \refs{\KKLMV,\KKV}
\eqn\rigidlimit{
\eqalign{
\exp\left(-{1\over g_s^2}\right)&=\exp(-t_1)= c_1\epsilon^4 \Lambda^4 \cr
t_2&=\epsilon c_2 a\ ,}}
If we choose $c_1={1\over 2}$ and $c_2=2$, $\Lambda$ and $a$
turn out to be those
of the Seiberg-Witten theory. We will make this choice in the
following.

Our aim is to solve the topological
$B$-model on this geometry at higher genus and to take the
field theory limit \rigidlimit\ of these contributions.
In sect.~3 we will derive the genus one result directly as the
holomorphic coupling of the ${\cal N}=2$ chiral multiplet to
gravity and in sect.~4 we will recover it from non-planar
contributions to the
free energy of the cubic matrix model, evaluated at the minimum
of the effective superpotential. It has been shown explicitly in
\DGKV\ that the effective superpotential is computed by the planar diagrams.
Our results thus provide a test of the Dijkgraaf-Vafa conjecture beyond
the planar approximation.

The leading structure of the period vector in the large complex
structure variables near $z_i\equiv e^{-\hat t_i}=0$\foot{A schematic view of
the moduli space can be found in \AKMV . } is
\eqn\periodP{
\Pi=(1,t_1=\log(z_1)+ \sigma,t_2=\log(z_2)+ \sigma ,F_q=t_1 t_2 +\rho)^t\,,}
where $\sigma=2(z_1 + z_2) + 3 (z_1^2 + 4 z_1 z_2 + z_2^2)+O(z^3)$ and
$\rho=4(z_1+z_2) +(9 z_1^2+32 z_1 z_2+ 9 z_2^2)+O(z^3)$. The formulae
in \CKYZ\ give the whole expansions. We denote $q_i=\exp(-t_i)$.
With $q=q_1/q_2$ and $Q=q_2$ we can obtain the
genus zero prepotential by integrating $F_q=- 2 q {\partial \over
\partial q}F^{(0)}(q,Q)$, cf. \CKYZ.  For later use we
define $Q=:e^{-T}$ and $q=:e^{-t}$.

The topological amplitudes $F^{(g)}$ for $g>0$ are not holomorphic in the complex
structure parameters, but various holomorphic limits exist. For example,
if we expand them around the large complex structure point where
$q_i\to 0$ one can take the
holomorphic limit $\bar q_i\to 0$ \BCOVone\ in which $F^{(g)}$ has the form
\eqn\genFg{
F^{(g)}=C^{(g)}+\sum_{m,n\geq0\atop(m,n)\neq(0,0}^{\infty}c_{nm}^{(g)}q_1^n q_2^m,}
where $C^{(g)}$ stands for classical terms. The expansion
parameters $c_{nm}^{(g)}$ are
the Gromov-Witten invariants for the maps of genus $g$ world-sheets with
bidegree $(m,n)$, while the
classical terms arise from maps of bidegree $(0,0)$. In our case we have
$C^{(0)}={1\over 24}t_1^3-{1\over 8} t_1^2 t_2-{1\over 8} t_1 t_2^2+{1\over 24} t_2^3$,
$C^{(1)}=-{t_1\over 24}-{t_2\over 24}$ and $C^{(g)}=(-1)^g
{|B_{2g} B_{2 g-1}| \over 4 g (2g-1) (2g-1)!}\chi$ ($\chi$ is the `regularized'
Euler number of the non-compact target space).

In the decoupling limit \rigidlimit\ the topological amplitudes
$F^{(g)}$ compute the holomorphic part of the low-energy effective
${\cal N}=2$ SUSY field theory. In particular the complexified
effective gauge coupling ($\tau={4\pi i \over g^2}+ {\theta \over
2 \pi}$) is obtained in the field theory limit as  $\tau(u)=
{\p^2\over\p a^2}{\cal F}^{(0)}=\lim_{\epsilon\rightarrow 0}
4\partial^2_{t_2} F^{(0)}$~\KKV, where the $\log(\epsilon)$ terms,
which appear in this limit, have been absorbed in the bare
coupling and $u(a)$ parameterizes the field theory moduli space.
Similarly, the holomorphic functions ${\cal F}^{(g)}(u(a))$, which
multiply the following combinations of the self-dual part of the
graviphoton field  strength, $F_+$, and of the curvature tensor,
$R_+$, in the effective Lagrangian \eqn\efffg{{\cal F}^{(g)}(u(a))
F_+^{2g-2} R_+^2\ } can be calculated from $F^{(g)}$ in the limit
\rigidlimit. On dimensional grounds, these terms are suppressed by
$M_{pl}^{(2-2g)}\sim\epsilon^{(2g-2)}$.

A non-vanishing $n$-space-time-instanton contribution to
${\cal F}^{(g)}(u)=\lim_{\epsilon\rightarrow 0}\epsilon^{(2g-2)}F^{(g)}$
arises from the infinite number of terms in $F^{(g)}$ at fixed
degree $n$ (i.e. w.r.t. the base) in \genFg.
This contribution must have the structure
$a^{2-2g}\left(\Lambda^4 \over a^4\right)^n$. This requires that
the corresponding term in $f^{(g)}$ scales like
${1\over (a \epsilon)^{2g-2}} \left(\Lambda^4 \over a^4\right)^n\sim
{q_1^{4n} \over (1-q_2)^{4n+2g-2}}$. This in turn implies that
$c^{(g)}_{nm}$ grows with $m$ as
\eqn\growrate{{\hat c}^{(g)}_{nm}=\gamma_n m^{4n-3+2g}.}
A similar argument was given in~\KKV\ for $g=0$. The
proportionality factor $\gamma_n$ is directly related to the
gauge theory space-time instantons. E.g. for the gauge coupling we have
\eqn\chat{
{\hat c}_{nm}^{(0)}={2^{(4n-1)}\over(4n-3)!n}{\cal F}_n m^{4n-3}\,,}
with ${\cal F}_n$ the $n$-instanton contribution
to ${\cal F}^{(0)}$ as calculated in \KLT.

The topological amplitudes for this model in the large volume
limit have been calculated for $g=0$ in \KKV\ and for $g=1$ in
\CKYZ. Generally, the amplitudes $F^{(g)}$ for $g>1$ are defined
recursively in terms of all amplitudes with lower $g$ and the
propagators $S^{ij}$ as defined in \BCOVtwo. It can be shown that
for this model the propagators can be chosen to vanish except for
$S^{tt}$ \AKMV. It can be determined from the $g=0,1$ amplitudes
by using the simplification of the holomorphic anomaly equation
which occurs at genus one in the case with only one non-zero
propagator:
\eqn\ftopi{\partial_{t} F^{(1)}= S^{tt} \partial^3_t F^{(0)} +\partial_{t}
\sum_{r} a_r \log(\Delta_r)\ . }
Here the sum is over the components of the discriminant loci which
are (i) are the conifold locus $\Delta_1=1-8(z_1+z_2)+16
(z_1^2+z_2^2)-32 z_1 z_2$ and (ii) the divisors $\Delta_2=z_1
z_2$  at large radius limit. The coefficients $a_r$ parameterize
the holomorphic ambiguity. A convenient choice is $a_1=-{1\over
12}$ and $a_2={1\over 12}$ which yields, using the genus $0,1$
results, the expansion of $S^{tt}$
\eqn\propstt{S^{tt}={1\over 2}+ 4 q_1 q_2+16 (q_1^2 q_2+q_1 q_2^2)+
40 (q_1^3 q_2+q_1 q_2^3)+188 q_1^2 q_2^2+{\cal O}(q^5)\ .}
With the recursive definition of the $F^{(g)}$ \BCOVtwo, worked
out for the local $B$-model in \KZ , the higher $F^{(g)}$  can be
calculated up to the holomorphic ambiguity. We have fixed the
latter from the knowledge of the absence of holomorphic curves of
low degree\KKVtwo.

We have calculated the Gromov-Witten invariants using the above
procedure and found agreement with genus 2 results of \Hosono, who
used the same method and at higher genus with the ones evaluated
in \AMV\ using Chern-Simons theory. In particular we find that the
$c^{(g)}_{nm}$ are
\eqn\cnm{
c_{nm}^{(g)}=P_n^{(4 n-3 + 2 g)}(m)=
p_n^{(2 n-2 + 2 g)}(x){\prod_{k=1}^{2n-1} (k+m)}}
where $P^{(d)}_n(x),p^{(d)}_n(x)$ are polynomials of degree $d$.
The remarkable fact that the Gromov-Witten invariants lie on
polynomials is a consequence of the embedding of the $SU(2)$
${\cal N}=2$ supersymmetric gauge theory in the string geometry
and holds therefore also for other geometries like ${\cal
O}(-K)\rightarrow \F_n$ with $n=1,2$, which allow for such an
embedding \KKV. We can give the topological string amplitudes
exactly to all orders in $q_2$ by specifying the $P_n^{(2 n-2 + 2
g)}(x)$ or alternatively by writing \eqn\growthbehaviour{
\eqalign{ F^{(0)}&=C^{(0)}(t)-2\Bigl({\rm Li}_{3}(q_2)
+\sum_{n=1}^\infty{q_1^n\over n^3(1-q_2)^{4n -2}}
h^{(0)}_n(q_2)\Bigr)\cr F^{(1)}&=C^{(1)}(t)-{1\over 6}\Bigl({\rm
Li}_{1}(q_2) -\sum_{n=1}^\infty {q_1^n\over n(1-q_2)^{4n}}
h^{(1)}_n(q_2)\Bigr)\cr F^{(g)}&=C^{(g)}+{(-1)^g B_{2 g}\over
g(2g-2)!} \sum_{n=0}^\infty{q_1^n\over n^{3-2g} (1-q_2)^{4 n +2
g-2}} h^{(g)}_n(q_2)\ .}}

The $h^{(g)}_n$ are polynomials of degree $2(n+g-1)$ whose
symmetric coefficients determine all orders in $q_2$  for given
a order in $q_1$. E.g. for $g=0$ we find
\eqn\hzeros{
\eqalign{
h^{(0)}_1=&1\cr
h^{(0)}_2=&1+18 q+q^2\cr
h^{(0)}_3=&1+98 q+450 q^2+98q^3 +q^4\cr
h^{(0)}_4=&1 + 306\,q + 4851\,{q^2} + 13188\,{q^3} + 4851\, q^4+\ldots\cr
h^{(0)}_5=&1 + 732\,q + 26903\,{q^2} + 206434\,{q^3} + 426060\,{q^4} +\ldots\cr
h^{(0)}_6=&1 + 1490\,q + 105315\,{q^2} + 1660604\,{q^3} +
  8358292\,{q^4} + 14651604\,{q^5} +\ldots \cr
         &\vdots }}
The missing terms, indicated by $\dots$, can be recovered
from the ones displayed via $h_n^{(g)}(1/q)=q^{-(2(n+g-1)}h^{(g)}(q)$.
\hzeros\ and \rigidlimit\ allow to determine the ${\cal F}_n$ and
also provides an expansion scheme for the string corrections. E.g.
using $c_1={1\over 2}$ and $c_2=2$ we expand $F^{(0)}$ in the
limit \rigidlimit , using ${\rm Li}_3(1-\epsilon
a)=\zeta(3)-{\pi^2\over 6} a\epsilon +( {3\over 4} -{\pi^2\over
12} - {1\over 2} \log(\epsilon a)) a^2 \epsilon^2+{\cal
O}(\epsilon^3)$. This yields to the relevant order, namely
$\epsilon^2$, from \growthbehaviour\ and \hzeros
\eqn\swfnull{{\cal F}^{(0)}={a^2}\Bigl(\log(x)+c^{(0)} -{x\over
2^5} -{5 x^2\over 2^{14}} -{3 x^3\over 2^{18}} -{1469 x^4\over
2^{31}} -{4471 x^5\over 5\cdot2^{34}} -{40397 x^6\over 2^{43}}
-{441325 x^7\over 7\cdot 2^{47}}+{\cal O}(x^8)\Bigl)\ ,} where
$x=\left(\Lambda\over a\right)^4$ and $c^{(0)}$ is related to
the bare gauge coupling into which a
$\log(\epsilon)$ term has been absorbed.
This matches the Seiberg-Witten prepotential \KLT .

The corresponding expressions for genus one are
\eqn\hones{\eqalign{
h^{(1)}_1&=(1-q)^2\cr
h^{(1)}_2&=1-2 q - 94 q^2- 2 q^3+q^4\cr
h^{(1)}_3&=1 - 1137\,{q^2} - 3872\,{q^3} - \ldots  \cr
h^{(1)}_4&= 1+ 4\,q - 6818\,{q^2} - 72168\,{q^3} - 158262\,{q^4} -\ldots \cr
h^{(1)}_5&=1+10\,q-28440\,{q^2}-643440\,{q^3}-3622665\,{q^4}-6479092\,{q^5}
-\ldots\cr
h^{(1)}_6&=1+18\,q-94008\,{q^2}-3827252\,{q^3}-41834673\,{q^4}- 167100606\,{q^5} -
265697392\,{q^6} -\ldots\cr
&\vdots}}
The holomorphic quantities of the gauge theory from higher ${\cal F}^{(g)}$ \efffg\  can be
extracted in the limit \rigidlimit\ at order $\epsilon^{2-2g}$,
as explained above. For ${\cal F}^{(1)}$ we obtain from
\growthbehaviour\ and \hones
\eqn\swfone{{\cal F}^{(1)}=-{1\over 24} \log(x)+c^{(1)}
+{x^2\over 2^{13}}
+{x^3\over 3\cdot 2^{14}}
+{1647 x^4\over 2^{29}}
+{981 x^5\over 2^{31}}
+{450137 x^6\over 3\cdot 2^{41}}
+{45111 x^7\over 2^{42}}+{\cal O}(x^8)\ ,}
this expression matches exactly the one calculated in the twisted
${\cal N}=2$ gauge theory in the next section and the matrix model
calculation in section 4.

We have calculated the $h^{(g)}_i$ with the topological 
B-model methods described
in \AKMV, albeit at another expansion point, up to genus $3$.
These expressions can be found in appendix A. In the gauge theory limit
we get the following expressions for the holomorphic functions in \efffg\
\eqn\swfhigher{\eqalign{\
{\cal F}^{(2)}&={1\over a^2}\Bigl(-{1\over 480}
-{11 x^2\over 2^{18}}
-{117 x^3\over 2^{22}}
-{171201 x^4\over 2^{34}}
-{1919923 x^5\over 5\cdot 2^{37}}
-{96877135 x^6\over 2^{47}}
+{\cal O}(x^7)\Bigr)\ , \cr
{\cal F}^{(3)}&={1\over a^4}\Bigl(
-{1\over 8064}
+{7 x^2\over 2^{19}}
+{293 x^3\over 2^{23}}
+{985823 x^4\over 2^{35}}
+{4069345 x^5\over 5\cdot 2^{38}}
+{416333277 x^6\over 2^{46}}
+{\cal O}(x^7)\Bigr)\ .}}

\subsec{The B-model on the local geometry $\widehat {II}$}

As explained before, the geometry $\widehat {II}$ describes the
low-energy dynamics of $U(N)$ super Yang-Mills theory with a cubic
tree-level superpotential, and Dijkgraaf and Vafa conjectured in
\DVone\ that the amplitudes of the 
topological B-model on this geometry, which
correspond to the holomorphic couplings of the gauge theory, is captured by
the cubic matrix model. At genus zero this was checked by
noticing that the planar solution of the matrix model reproduces
the special geometry of $\widehat{II}$ \DVone, and it
was also confirmed in \DGKV\ by using matrix model perturbation
theory. To check this at genus one, we shall calculate $F^{(1)}$.
As explained in \refs{\BCOVone,\BCOVtwo}, $F^{(1)}$ is a section of a
determinant line bundle over the complex moduli space, which was
studied by Ray and Singer. $F^{(1)}$ satisfies a holomorphic
anomaly equation which can be readily integrated up to a
holomorphic ambiguity to be discussed below.

To proceed
we need the periods $S_1,S_2$ from \CIV, whose notation we adopt.
It is convenient to change variables
$(a_1^-,a_1^+,a_2^-,a_2^+)\equiv(x_1,x_2,x_3,x_4)
\to(\Delta_{21},\Delta_{43},Q,I)$
where
\eqn\variables{
\eqalign{\Delta_{21}&\equiv{1\over2}(x_2-x_1)\,,\qquad
\Delta_{43}\equiv{1\over2}(x_4-x_3)\cr
Q&\equiv{1\over2}(x_1+x_2+x_3+x_4)=-{m\over g}\cr
I&\equiv{1\over2}[(x_3+x_4)-(x_1+x_2)]=
\sqrt{\Delta^2-2(\Delta_{21}^2+\Delta_{43}^2)}}}
We will also use $z_1=\Delta^2_{43}$ and $z_2=\Delta^2_{21}$ and
$\Delta=(a_1-a_2)={m\over g}$.

The periods $S_i(z_1,z_2,g,\Delta)$ were computed in \CIV
\foot{Note that the sums in the expressions for the periods
$S_1$ and $\Pi_1$
in appendix B of \CIV\ should read $\sum_{n=1}^\infty c_n \Delta_{21}^{2n}
{F^{2n-2}(0)\over (2n-2)!}$ and  $\sum_{n=1}^\infty c_n \Delta_{21}^{2n}
{G^{2n-2}(0)\over (2n-2)!}$, respectively.}.
To express the $B$-model amplitudes in terms of the $S_i$ we need
the inverse relations
\eqn\SDelta{\eqalign{
z_1(S_1,S_2,g,\Delta)&={4\over g \Delta}S_1+
{8\over g^2\Delta^4}(2 S_1^2-3 S_1S_2)+{8\over g^3\Delta^7}S_1(5 S_1-13 S_2)
(4 S_1-3 S_2)+{\cal O}(S^4)\cr
z_2(S_1,S_2,g,\Delta)&=z_1(S_2,S_1,g,-\Delta)}}
According to \BCOVone\ and \BCOVtwo\ and taking the simplification
in the local case \KZ\ into
account, we expect the holomorphic $\bar S_i\rightarrow 0$ limit
of the genus one B-model amplitude to be,
up to an additive constant,
\eqn\Fonetop{
F^{(1)}={1\over 2}\log\left(\det\left(\partial z_i\over\partial S_j\right)
f(z)\right)\,.}
For the holomorphic ambiguity $f(z)$ we make the Ansatz
$f(z)=(\prod_i\Delta_i^{\kappa_i}) I^{\kappa_3}$. The $\Delta_i$ are
components of the discriminant of the Riemann surface. It splits
into two factors, the conifold divisor $\Delta_1=z_1\cdot z_2$ and
$\Delta_2=(x_3-x_1)(x_3-x_2)(x_4-x_2)(x_4-x_1)$. The exponents
$\kappa_i$ determine the leading behavior of $F^{(1)}$ at the
degeneration loci. At the conifold we expect \Vafa\
$\kappa_1=-{1\over6}$. We fixed the $\kappa_i$ by comparing with
any three coefficients of the matrix model computation, which we will
present in sect.~4, to $\kappa_1=-{1\over 6}$, $\kappa_2={2\over3}$ and $\kappa_3=-1$. 
Note that $I=0$ is not a discriminant component 
of the local geometry. The above 
ansatz for $f$ was motivated by simplicity. $I$ seems to appear in $f$ as 
a branch locus in the moduli space similarly as the $Z_5$ orbifold 
point $\Psi=0$ appears in the ambiguity of $F^{(1)}$ for the 
quintic in \BCOVtwo. 
$F^{(1)}$ has the expansion
\eqn\FonetypeII{
\eqalign{
F^{(1)}=&-{1\over12}\log\Bigl({S_1 S_2\over\Lambda^6}\Bigr)+{1\over 6}(S_1-S_2)
+{g^4\over m^6}\left(\frac{7}{3}S_1^2-\frac{31}{3}S_1 S_2
+\frac{7}{3}S_2^2\right)\cr
&+{g^6\over m^9}\left(\frac{332}{9}S_1^3
-\frac{923}{3}S_1^2 S_2+\ldots\right)\cr
&+{g^8\over m^{12}}\left(\frac{1864}{3}S_1^4-\frac{47083}{6}S_1^3 S_2
+15349{S_1}^2{S_2}^2 -\dots \right)\cr
&+{g^{10}\over m^{15}}\left( \frac{54416}{5}S_1^5 - 187528{S_1}^4 S_2
+ 570066{S_1}^3{S_2}^2 - \ldots \right)\cr
&+{g^{12}\over m^{18}}\left(\frac{1762048}{9}S_1^6
-\frac{12980560{S_1}}{3}S_1^5 S_2+\frac{54863776}{3}S_1^4 S_2^2 -
\frac{256344964}{9}S_1^3 S_2^3 +\ldots\right)\cr &+ {\cal O}(S^7),}}
where $\dots$ means antisymmetric completion.

\newsec{${\cal F}^{(1)}$ from topological field theory}

In the decoupling limit \rigidlimit, the amplitudes 
${\cal F}^{(g)}$  give corrections
to the low energy effective action of ${\cal N}=2$ super Yang-Mills theory
of the form \efffg. In particular,
${\cal F}^{(1)}$ gives a term
\eqn\effone{
\int d^4 x {\cal F}^{(1)} (a) {\rm Tr} R_+^2}
which can be integrated to obtain
\eqn\coupl{ {1 \over 2}{\cal F}^{(1)}(a) \Bigl( \chi -{3 \over 2}
\sigma \Bigr). }
Here $a$ is a coordinate in the moduli space of
${\cal N}=2$ super Yang-Mills, and $\chi$ and $\sigma$ denote the
Euler characteristic and signature of the four-manifold,
respectively. A coupling to gravity like \coupl\ is natural when
the theory is embedded in string theory, as we are doing here.
Another way in which couplings to gravity are relevant is when the
${\cal N}=2$ theory is topologically twisted. Indeed, as shown in
\Witten, the low-energy effective action of twisted ${\cal N}=2$
super Yang-Mills on an arbitrary four-manifold contains terms
proportional to the Euler characteristic and the signature (these
terms appear in any twisted gauge theory with ${\cal N}=2$
\refs{\MW,\others} or ${\cal N}=4$ \VW\ supersymmetry). As pointed
out in \refs{\VW,\DVtwo}, when the manifold is hyperK\"ahler, the
twisted and the physical theory agree, and we should therefore be
able to compare the coupling \coupl\ with the corresponding result
in topological field theory. This was done in \DVtwo\ for the
${\cal N}=4$ theory.

The geometry $I$ of the preceding section engineers ${\cal N}=2$ Yang-Mills
theory with gauge group $SU(2)$ in the limit \rigidlimit.
On the other hand, the coupling to gravity that appears in the
effective action of the twisted theory can
be written \refs{\Witten,\MW} in terms of data of the Seiberg-Witten
elliptic curve \refs{\SW,\SWtwo}:
\eqn\tft{
-{\chi \over 2}\log\Bigl({\Lambda da\over du}\Bigr)
+{\sigma\over 8}\log\Bigl({\Delta\over\Lambda^4}\Bigr)\,,}
where $a(u)$ is the $a$-period of the Seiberg-Witten differential,
$u$ the coordinate on the moduli space, and $\Delta$ is the
discriminant of the curve. This should agree with \coupl\ on a
compact four-dimensional hyperK\"ahler manifold. There are two
such manifolds: $T^4$ and $K3$. For $T^4$, $\chi=\sigma=0$, and
both \coupl\ and \tft\ are trivial. For K3, $\chi=24$,
$\sigma=-16$, and by comparing \coupl\ to \tft\ we find:
\eqn\fI{
{\cal F}^{(1)}(a)=-{1\over 2}\log\Bigl({\Lambda da\over du}\Bigr)
-{1\over 12} \log\Bigl({\Delta\over\Lambda^4}\Bigr)\,.}
Using the explicit expressions $\Delta= u^2 -\Lambda^4$,
and the explicit relation between $u$ and $a$ worked out in \KLT,
\eqn\uarel{
u(a) = 2 a^2 \biggl(1 + {1 \over 2^4}\Bigl({\Lambda\over a}\Bigr)^4 +
 + {5 \over 2^{12}}\Bigl({\Lambda\over a}\Bigr)^8 +
{ 9  \over 2^{17}}\Bigl({\Lambda\over a}\Bigr)^{12} + {1469  \over
2^{28}}\Bigl({\Lambda\over a}\Bigr)^{16} + \cdots \biggr),}
we computed the first few non-trivial corrections:
\eqn\ftop{ \eqalign{ {\cal F}^{(1)}(a)=&{1\over 6}\log\Bigl( 32
{a\over \Lambda} \Bigr) +{1\over 2^{13}}\Bigl({\Lambda\over
a}\Bigr)^8+{1\over 3\cdot 2^{14}}
 \Bigl({\Lambda \over a}\Bigr)^{12}
+{1647 \over 2^{29}}\Bigl({\Lambda \over a}\Bigr)^{16} \cr
\noalign{\vskip.2cm}
&\qquad +{981 \over 2^{31}}\Bigl({\Lambda \over a}\Bigr)^{20} +
{450137 \over 3\cdot 2^{41}}\Bigl({\Lambda \over a}\Bigr)^{24} +
{45111 \over 2^{42}}\Bigl({\Lambda \over a}\Bigr)^{28}+ \cdots \cr }}
This is precisely what we found in sect.~2.

It is interesting to notice that, when ${\cal F}^{(1)}$ is written
in terms of modular forms (by using for example the results in
\MW), one finds
\eqn\foneta{ {\cal F}^{(1)}=-\log \eta (\tau).}
In this equation $\tau$ is the modular parameter of the
Seiberg-Witten curve when written in the $\Gamma(2)$ description
of \SW, and it is related to the modular parameter $\tau_0$ of the
$\Gamma^0(4)$ description of \refs{\SWtwo,\KLT} by $\tau =\tau_0/2$.
Equation \foneta\ is in agreement with the general considerations
in \DVthree\ relating $F^{(1)}$ to $\eta(\tau)$ whenever the local
Calabi-Yau geometry reduces to a Riemann surface.

\newsec{Matrix model considerations}

The conjecture of Dijkgraaf and Vafa states that the cubic matrix model
captures all the information about the $F^{(g)}$ of the
$\widehat{II}$ geometry. On the other hand, one can consider a
limit that recovers the Seiberg-Witten exact solution to $SU(2)$,
${\cal N}=2$ super Yang-Mills theory, together with its
gravitational corrections \refs{\CV,\DVthree,\DGKV}, which are
given by the non-planar sector of the matrix model. In this section
we give some results for the cubic matrix model beyond the planar
approximation. We compute the free energy up to sixth order in
perturbation theory, providing in this way information up to
$g=3$, and we use the loop equations to find an exact expression
for $F^{(1)}$.

\subsec{Perturbative calculation}

The cubic matrix model has the potential
\eqn\cubic{
W(\Phi)=\tr\Bigl({m\over 2} \Phi^2 + {g \over 3} \Phi^3\Bigr)
=\sum_{i=1}^N\Bigl({m\over2}\l_i^2+{g\over3}\l_i^3\Bigr).}
Classically the potential is extremized if the eigenvalues are in
either one of the two critical points, i.e. at $a_1=0$ or
$a_2=-m/g$. Quantum-mechanically the eigenvalues form two bands
and there is eigenvalue tunneling between the two bands. We want
to consider the metastable vacuum in which $N_1$ eigenvalues are
at $0$ and $N_2$ eigenvalues are at $-m/g$, with $N_1$ and $N_2$
fixed, subject only to the condition that $N_1+N_2=N$. This
corresponds, in matrix model terminology, to a two-cut solution.
This issue has already been addressed recently \DGKV. Here we will
use a slightly different approach which avoids introducing ghost
degrees of freedom. While the authors of \DGKV\ considered only
the planar limit of the matrix model, we also compute non-planar
contributions. While the former contain the information about the
effective superpotential, the holomorphic couplings of the field
theory to gravity are obtained from the non-planar part of the
free energy.

The partition function $Z$ and free energy $F$ of the matrix model are:
\eqn\eigenz{
Z=e^{F}={1\over {\rm Vol}(U(N))}\int D\Phi\,\Phi e^{-W(\Phi)}=
{1\over N!(2\pi)^N}\int\prod_i d\lambda_i\Delta^2(\lambda){\rm e}^{-{m\over 2}
\sum_i \lambda_i^2 -{g\over 3} \sum_i \lambda_i^3},}
where $\Delta(\lambda)=\prod_{i<j} (\lambda_i-\lambda_j)$
is the Vandermonde determinant.
We expand around the vacuum with $\l_i=0$ for $i=1,\dots, N_1$ and
$\l_i=-{m\over g}$ for $i=N_1+1,\dots, N$. Denoting the fluctuations by
$\mu_i$ and $\nu_j$, the Vandermonde determinant becomes
\eqn\vander{
\Delta^2(\lambda)=\prod_{1\le i_1<i_2\le N_1}\!\!\!\! (\mu_{i_1}-\mu_{i_2})^2
\prod_{1\le j_1<j_2 \le N_2}\!\!\!\!(\nu_{j_1}-\nu_{j_2})^2\,
\prod_{{1\le i\le N_1\atop 1\le j\le N_2}}
\Bigl(\mu_{i}-\nu_j + {m\over g}\Bigr)^2.}
We also expand the potential around this vacuum and get
\eqn\action{
W=\sum_{i=1}^{N_1}\left({m\over2}\mu_i^2+{g\over 3}\mu_i^3\right)
-\sum_{i=1}^{N_2}\left({m\over2}\nu_i^2-{g\over 3}\nu_i^3\right)
+{m^3 \over 6 g^2}N_2.}
Notice that the propagator of the fluctuations around
$-m/g$ has the `wrong' sign.
The interaction between the two sets of eigenvalues,
which is given by the last factor in \vander,
can be exponentiated and included in the action, as in \AKMV. This
generates an interaction term between the two eigenvalue bands
\eqn\effe{
W_{\rm int}=-2N_1 N_2\log{m\over g}+2\sum_{k=1}^{\infty}{1\over k}
({g\over m})^k\sum_{i,j}\sum_{p=0}^k  (-1)^p {k \choose p}
\mu_i^{p} \nu_j^{k-p}.}
By rewriting the partition function in terms of matrices instead of
their eigenvalues, we can
represent this model as an effective two-matrix model,
involving an $N_1 \times N_1$
matrix $\Phi_1$, and an $N_2\times N_2$ matrix $\Phi_2$:
\eqn\efftwo{
Z={1\over {\rm Vol}(U(N_1)) \times {\rm Vol}(U(N_2))}
\int D\Phi_1 D\Phi_2 {\rm e}^{-W_1(\Phi_1)-W_2(\Phi_2)
-W(\Phi_1, \Phi_2)},}
where
\eqn\terms{
\eqalign{
W_1 (\Phi_1)=&+\tr\Bigl({1\over 2} m \Phi_1^2 + {g \over 3} \Phi_1^3\Bigr), \cr
W_2 (\Phi_2)=&-\tr\Bigl({1\over 2} m \Phi_2^2 - {g \over 3} \Phi_2^3\Bigr), \cr
W_{\rm int}(\Phi_1,\Phi_2)=&~2\sum_{k=1}^{\infty}{1\over k}({g\over m})^k
\sum_{p=0}^k (-1)^p{k\choose p}{\rm tr}\,\Phi_1^p~
{\rm tr}\,\Phi_2^{k-p}\cr
&\qquad\quad+N_2 W(a_2)+N_1 W(a_1)+2 N_1 N_2\ln\Bigr({m\over g}\Bigr).}}
where $\tr\,\Phi_1^0=N_1$, $\tr\,\Phi_2^0=N_2$, $W(a_1)=0$ and
$W(a_2)={m^3\over 6 g^2}$.
We have dropped the statistical factor
$N!/(N_1! N_2!)$ which counts the number of ways to distribute the
$N$ eigenvalues among the two critical points of the potential.
This two-matrix model is (perturbatively) well defined if we choose
$\Phi_1$ hermitian and $\Phi_2$ anti-hermitian, i.e.
$\mu_i$ real and $\nu_j$ imaginary.
We can now compute the free energy $F=\log(Z)$ is a straightforward way.
$F$ consists of two parts, a perturbative part
$F_{\rm pert}=\log\bigl(Z(g)/Z(g\!\!=\!\!0)\bigr)$
which vanishes for the free (Gaussian) model
and a non-perturbative part
$F_{\rm n.p.}$ which gets contributions from the $U(N)$ group-volume factor.
Both can be computed in a straightforward way. The perturbative part can
be expanded as
\eqn\Fpertdef{
F_{\rm pert}=-N_1 W(a_1)-N_2 W(a_2)+2 N_1 N_2 \ln\Bigl({m\over g}\Bigr)+
\sum_{h=1}^\infty\sum_{g\ge0\atop h+2-2g>0}
\Big({g^2\over m^3}\Big)^h F_{h,g}(N_1,N_2)}
where
$F_{h,g}$ is a homogeneous polynomial in $N_1$ and $N_2$
of degree $h+2-2g$.
One finds, up to $h=6$,
\ninepoint
\eqn\Fpertmatrix{
\eqalign{
F_{\rm pert}&=-N_1 W(a_1)-N_2 W(a_2)+2 N_1 N_2\ln({m\over g})\cr
+&{g^2\over m^3}\left[\Bigl({2\over3}N_1^3-5 N_1^2 N_2+5 N_1 N_2^2
-{2\over3}N_2^3\Bigr)+{1\over 6}(N_1-N_2)\right]\cr
+&{g^4\over m^6}\left[\Bigl({8\over3}N_1^4-{91\over3}N_1^3 N_2+59 N_1^2 N_2^2
-{91\over3}N_1 N_2^3+{8\over3}N_2^4\Bigr)
+\Bigl({7\over3}N_1^2-{31\over3}N_1 N_2+{7\over3}N_2^2\Bigr)\right]\cr
+&{g^6\over m^9}\left[\Bigl({56\over3}N_1^5-{871\over3}N_1^4 N_2
+{2636\over3}N_1^3 N_2^2-{2636\over 3}N_1^2 N_2^3+{871\over3}N_1 N_2^4
-{56\over3}N_2^5\Bigr)\right.\cr
&\qquad\quad\left.+\Bigl({332\over9}N_1^3-{923\over3}N_1^2 N_2
+{923\over3}N_1 N_2^2-{332\over9}N_2^3\Bigr)+{35\over6}(N_1-N_2)\right]\cr
+&{g^8\over m^{12}}\left[\Bigl({512\over3}N_1^6-{6823\over2}N_1^5 N_2
+{28765\over2}N_1^4 N_2^2-{67310\over3}N_1^3 N_2^3\pm\dots\Bigr)\right.\cr
&\qquad\quad\left.+\Bigl({1864\over3}N_1^4-{47083\over6}N_1^3 N_2
+15349 N_1^2 N_2^2\mp\dots\Bigr)
+\Bigl(338 N_1^2-1632 N_1 N_2+338 N_2^2\Bigr)\right]\cr
+&{g^{10}\over m^{15}}\left[
{9152\over 5}\Bigl(N_1^7-45118 N_1^6 N_2+247980 N_1^5 N_2^2
-540378 N_1^4 N_2^3\pm\dots\Bigr)\right.\cr
&\left.\qquad\quad
+\Bigl({54416\over5}N_1^5-187528 N_1^4 N_2+570066 N_1^3 N_2^2\mp\Bigr)
+\Bigl({66132\over5}N_1^3-120880 N_1^2 N_2\mp\dots\Bigr)\right.\cr
&\qquad\quad\left.+{5005\over3}(N_1-N_2)\right]\cr
+&{g^{12}\over m^{18}}\left[
\Bigl({65536\over 3}N_1^8-{1933906\over3}N_1^7 N_2
+{13258178\over3}N_1^6 N_2^2-{37761034\over3}N_1^5 N_2^3
+{52780010\over3}N_1^4 N_2^4\mp\dots\Bigr)\right.\cr
&\qquad\quad\left.+\Bigl({1762048\over9}N_1^6-{12980560\over3}N_1^5 N_2
+{54863776\over3}N_1^4 N_2^2
-{256344964\over9}N_1^3 N_2^3\pm\dots\Bigr)\right.\cr
&\qquad\quad\left.+\Bigl({1305280\over3}N_1^4-{18059582\over3}N_1^3 N_2
+11824166 N_1^2 N_2^2\mp\dots\Bigr)\right.\cr
&\qquad\quad\left.
+\Bigl({1680704\over9}(N_1^2+N_2^2)-{8748896\over9}N_1 N_2\Bigr)\right]
+\dots}}
\tenpoint
where $\dots$ means again anti-symmetric completion.
The non-perturbative contribution to the free energy is
\eqn\Fnonpa{
e^{F_{\rm n.p.}}=
m^{-{1\over2}(N_1^2+N_2^2)}(2\pi)^{-{1\over2}(N_1+N_2)}
\prod_{k=1}^{N_1}\Gamma(k)\,\prod_{l=1}^{N_2}\Gamma(l)\,.}
With the help of the asymptotic expansion (see e.g.\OV)
\eqn\asympt{
\ln\Big(\prod_{k=1}^N\Gamma(k)\Big)=
{N^2\over2}\ln N-{1\over 12}\ln N
-{3\over4}N^2
+{1\over2}N\ln 2\pi+\zeta'(-1)
+\sum_{g=2}^\infty{B_{2g}\over 4g(g-1)}{1\over N^{2g-2}}}
this becomes
\eqn\Fnp{\eqalign{
F_{\rm n.p.}&=
{1\over2}N_1^2\ln\Big({N_1\over m}\Big)
+{1\over2}N_2^2\ln\Big({N_2\over m}\Big)-{3\over4}(N_1^2+N_2^2)
-{1\over12}\ln(N_1 N_2)\cr
\noalign{\vskip.2cm}
&\qquad+2\zeta'(-1)
+\sum_{g=2}^\infty{B_{2g}\over 4g(g-1)}\Bigl({1\over N_1^{2g-2}}
+{1\over N_2^{2g-2}}\Bigr)}}
Comparing with the prepotential of the gauge theory with cubic
superpotential, which can be extracted from the results given in
\CIV, we find it to be in precise agreement with the leading
terms, at each order in the coupling constant, of the free energy
computed from the matrix model if we identify $N_i\to S_i$.
The only part of the gauge theory result which is not
directly determined by the matrix model is the dependence on
the cut-off $\Lambda$.

\subsec{$F^{(1)}$ from the loop equations}

The expression for the free energy \Fpertmatrix\ was obtained by
doing perturbation theory. However one can obtain analytic
expressions for the genus $g$ free energies by using the loop
equations of the matrix model. These equations were used
systematically in \ACKM\ in the one-cut case, and extended to the
multi-cut case in \Akemann. Before moving on to the comparison
with topological string theory calculations, we derive an analytic
expression for $F^{(1)}$ for the cubic matrix model which can be
easily expanded to high powers in $N_1$ and $N_2$. The computation
closely parallels the derivation in \Akemann\ of the two-cut
solution, to which we refer for further details. The reason why we
cannot directly adopt the result of \Akemann\ is that instead of
imposing the absence of eigenvalue tunneling between the two
bands (equal chemical potential), we fix $N_1$ and $N_2$, {\it
i.e.} we impose
\eqn\condition{
\int_{x_3}^{x_4}\rho(x) dx=N_1\,,\qquad
\hbox{and}\qquad
\int_{x_1}^{x_2}\rho(x) dx=N_2\,.}
instead of $\int_{x_2}^{x_3}\rho(x) dx=0$. In other words, we are
not looking for the true vacuum but consider a meta-stable vacuum
with fixed $N_1$ and $N_2$\foot{The complete matrix-model
partition function involves a sum over all possible eigenvalue
distributions and does not have a topological expansion, as
explained in \BDE. This subtlety is, however, not relevant here as
we fix $N_1$ and $N_2$.}. Here $\rho(x)$ is the eigenvalue density
which is given by the discontinuity of the resolvent of the matrix
$\Phi$. For the cubic superpotential
$W(x)={m\over2}x^2+{g\over3}x^3$ it is $\rho(x)={g\over2\pi
i}\sqrt{\prod_{i=1}^4(x-x_i)} ={1\over 2\pi
i}\sqrt{W'^2(x)-f_1(x)}$ where $f_1$ is a polynomial of order one
whose two coefficients parameterize the widths of the two branch
cuts. Note that $N_i$ are given by the same integrals as $S_i$ in
the gauge theory. The two conditions \condition\ are not
independent since $N_1+N_2=N$. The other conditions, which follow
from the asymptotic behavior of the resolvent, are exactly as in
\Akemann. Following the steps in \Akemann\ one finds (we give a
few intermediate results of the computation in appendix B)
\eqn\Fonematrix{
F^{(1)}=-{1\over24}\sum_{i=1}^4\ln M_i-{1\over2}\ln K(k)
-{1\over12}\sum_{i<j}\ln(x_i-x_j)^2
+{1\over8}\ln(x_1-x_3)^2+{1\over8}\ln(x_2-x_4)^2+{\rm const.}}
where $K(k)$ is a complete elliptic integral with modulus
\eqn\modulus{
k^2={(x_1-x_2)(x_3-x_4)\over(x_1-x_3)(x_2-x_4)}}
and the moments $M_i$ are defined as
\eqn\defmoments{
M_i=\oint_{\cal C}{dx\over 2\pi i}
{W'(x)\over (x-x_i)\sqrt{\prod_{i=1}^4(x-x_i)}}}
where the contour ${\cal C}$ encloses both cuts. For
cubic $W$, $M_i=g$ for all $i=1,\dots, 4$, as one shows
by deforming ${\cal C}$ such as to enclose infinity.

It remains to express $F^{(1)}$ as a series in $N_1$ and $N_2$,
with coefficients depending on $m$ and $g$, the two parameters of
$W$. For this purpose it is convenient to change variables, as in
\CIV, $(x_1,x_2,x_3,x_4)\to(\Delta_{21},\Delta_{43},Q,I)$, where
the explicit relations were given in \variables.
Inserting \SDelta\ with $S_i$ replaced by $N_i$, we obtain
\eqn\Foneexpansion{ \eqalign{ F^{(1)}&=-\frac{1}{12}\log(N_1 N_2)+
{g^2\over m^3}\left(\frac{N_1}{6}-\frac{N_2}{6} \right) +{g^4\over
m^6}\left(\frac{7N_1^2}{3}-\frac{31N_1N_2}{3}
+\frac{7N_2^2}{3}\right)\cr \noalign{\vskip.2cm} &+{g^6\over
m^9}\left(\frac{332N_1^3}{9}-\frac{923N_1^2N_2}{3}\pm\ldots\right)
+{g^8\over m^{12}}\left(\frac{1864N_1^4}{3}
-\frac{47083N_1^3N_2}{6}+15349{N_1}^2{N_2}^2\mp\dots\right)\cr
\noalign{\vskip.2cm} &+{g^{10}\over
m^{15}}\left(\frac{54416{N_1}^5}{5}- 187528{N_1}^4N_2 +
570066{N_1}^3{N_2}^2\mp\ldots\right)\cr \noalign{\vskip.2cm}
&+{g^{12}\over m^{18}}\left(\frac{1762048{N_1}^6}{9}
-\frac{12980560{N_1}^5N_2}{3}+\frac{54863776{N_1}^4{N_2}^2}{3} -
\frac{256344964{N_1}^3{N_2}^3}{9}\pm\ldots\right)\cr &+{\cal
O}(N^7)}}
where $\ldots$ means, as before, antisymmetric completion. This
expression agrees exactly with the expansion of $F^{(1)}$ for the
local geometry II \FonetypeII\ that we obtained in section 2, with
the identification $S_i=N_i$, $i=1,2$. This provides a one-loop
test of the relation conjectured in \DVone.

Using the iterative procedure developed in \ACKM, one can,
with some effort, derive expressions for the higher
genus contributions to the free energy.

\subsec{ ${\cal N}=2$ Yang-Mills from the
matrix model }


As explained in \CV\DVthree\DGKV, using the results of the cubic
matrix model one can derive results for $SU(2)$, ${\cal N}=2$
Yang-Mills theory. The idea is the following: the cubic matrix
model corresponds to the ${\cal N}=2$ theory broken down to ${\cal
N}=1$ by adding the tree level superpotential $W(\Phi)$ for the
chiral superfield $\Phi$. By taking $g, m \rightarrow 0$ and
keeping $\Delta$ fixed, we recover the pure ${\cal N}=2$ theory.
In general, a tree level superpotential of order $r+1$ allows one
to recover the moduli space of the $SU(r)$ theory. In our case, we
can recover the $SU(2)$ theory, and $\Delta$ will then be related
to the well-known $u$-modulus of Seiberg-Witten theory \SW\SWtwo.
Some results for the prepotential for general $r$ have been
recently obtained in \NSW.

In order to recover the ${\cal N}=2$ results, we first have to
find the relation between the matrix model variables and the usual
variables of the Seiberg-Witten solution. Following \DGKV, we
extremize the effective superpotential of the gauge theory $W_{\rm
eff}={\p F^{(0)}\over\p S_1}+{\p F^{(0)}\over \p S_2}$ where (cf.
\Fpertmatrix\ and \Fnp\ with $N_i\to S_i$ and the $\Lambda$
dependence added on dimensional grounds)
\eqn\Fzero{\eqalign{ F^{(0)}&={1\over2}S_1^2\log\Bigl({S_1\over
m\Lambda^2}\Bigr) +{1\over2}S_2^2\log\Bigl({S_2\over
m\Lambda^2}\Bigr) -{3\over4}(S_1^2+S_2^2) +2S_1 S_2\log\Bigl({m\over
\Lambda g}\Bigr)\cr &\qquad +{1\over g\Delta^3}\Bigl({2\over3}S_1^3-5
S_1^2S_2+5 S_1 S_2^2 -{2\over3}S_2^3\Bigr)+{\cal O}(S^4).}}
Here we have ignored terms ${\cal O(S)}$ since they will not contribute
to the positions of the extrema of $W_{\rm eff}$.
Extremizing $W_{\rm eff}$,
{\it i.e.} solving $\p_{S_i}W_{\rm eff}=0$, requires $S_1=-S_2\equiv S$ with
\eqn\Smin{
{S\over g \Delta^3}=
\Bigl( {\Lambda \over \Delta} \Bigr)^4 \biggl( 1 +
6 \Bigl( {\Lambda \over \Delta} \Bigr)^4 + 140 \Bigl( {\Lambda \over
\Delta} \Bigr)^8 +4620\Bigl({\Lambda\over\Delta}\Bigr)^{12}+
180180\Bigl({\Lambda\over\Delta}\Bigr)^{16}+
7759752\Bigl({\Lambda\over\Delta}\Bigr)^{20}+
\cdots \biggr),}
as obtained in \DGKV. Note that the relation $S_1+S_2=0$ implies
that $f_1$ in \transciv\ vanishes. This, on the other hand means
that the contours which define $S_1$ and $S_2$ can be deformed
into each other without picking up a contribution from the point
at infinity. In other words, we can add this point and consider a
compact Riemann surface, as in Seiberg-Witten theory. However, the
scale $\Lambda$ which appears in \Smin\ is, a priori, not
identical with the scale appearing in Seiberg-Witten theory. To
find their relation, we relate the curve
$y^2=W'(x)^2+f_0=g^2\prod(x-x_i)$ to the Seiberg-Witten curve.
First we shift $x\to x-{1\over2}\Delta$. Using the relations
\variables\ and the solutions $\Delta_{21}(S,-S)$ and
$\Delta_{43}(S,-S)$ (c.f. \SDelta) one finds, after rescaling $y\to
gy$, $y^2=\Bigl(x^2-{1\over4}\Delta^2\Bigr)^2-4\Lambda^4$.
Comparing this with the $SU(2)$ SW curve,
$y^2=(x^2-u)^2-\Lambda_{\rm SW}^4$ leads to the identifications
\eqn\identifications{
u={1\over 4}\Delta^2\,,\qquad{\rm and}\qquad \Lambda={1\over\sqrt{2}}
\Lambda_{\rm SW}\,.}
This, together with \Smin, gives the relation between the variables in the
matrix model computation and the usual variables in Seiberg-Witten theory.

Since in order to obtain pure $SU(2)$ super Yang-Mills we take a
limit of the matrix model, in order to compare the results of the
matrix model calculation with the results obtained in sections 2
and 3 we have to be very careful and look for quantities that do
not vanish or diverge as $g \rightarrow 0$. For example, $S$
vanishes as $g \rightarrow 0$, while $S/g$ is independent of $g$
and can be expressed solely in terms of $\Delta$, as it is
apparent from \Smin. When we express $F^{(r)}$ in terms of $S/g$,
we see that it is given by $g^{2-2r}$, times a function of $S/g$.
Therefore, if $r\not=1$, the resulting quantity depends on $g$ and
vanishes or diverges as $g \rightarrow 0$. This is related to the
fact that $F^{(r)}$ is not a function in moduli space, but rather
a section of a line bundle ${\cal L}^{2-2r}$, {\it i.e.} for
$r\not = 1$ it is not invariant under the gauge transformations of
special geometry. This also indicates that, in order to compare
the matrix model calculations with the results of section 2 and 3,
we should focus on gauge-invariant quantities.

Following this idea, in the case of $F^{(0)}$ it is clear that the
appropriate quantities are the second derivatives w.r.t. the moduli,
{\it i.e.} the $\tau_{ij}$ couplings, which are independent of $g$ \CV.
One can check \DGKV\ that the $\tau_{ij}$ computed in the
matrix model lead to the right result for $\tau$ in the
${\cal N}=2$ theory. Since $F^{(1)}$ is gauge-invariant and can be
expressed in terms of $\Delta$, with no $g$ dependence, one should be able
to compare directly the $F^{(1)}$ of \ftop\ with the $F^{(1)}$ obtained in
the matrix model. Indeed, using \Smin, \identifications, and
the relation between the $u$ modulus and the $a$ variable given in \uarel,
we find that the $F^{(1)}$ of \Foneexpansion\ reproduces \ftop\ after
evaluating it at the extremum \foot{Note that the scale appearing in
sect.~3 is, in fact, $\Lambda_{\rm SW}$.}.
It seems less straightforward to relate the matrix model results for
$F^{(g)}$ for $g>1$ to the couplings ${\cal F}^{(g)}$  in Seiberg-Witten 
theory derived from string theory. In the final section we will, however, 
find agreement between the expressions \swfhigher\ and a 
recent conjecture by Nekrasov. 

\newsec{Comparison with instanton computations}

The semiclassical expansion of the ${\cal N}=2$ prepotential of
Yang-Mills theory can be obtained by direct computation of
instanton corrections. In a recent {\it tour de force} Nekrasov
\Nekrasov (see also \Flume) was able to provide general
expressions for the $n$-th
instanton contribution to the ${\cal N}=2$, $SU(N)$ prepotential,
with or without matter. The answer for the $n$-th
instanton correction has the form ${\cal F}_n (a, \epsilon_1,
\epsilon_2)$ and it is an analytic function in $\epsilon_{1,2}$.
For $\epsilon_1=-\epsilon_2=\epsilon$, one can expand this as
\eqn\expfn{ \epsilon^{-2} {\cal F}_n (a,
\epsilon)=\sum_{k=0}^{\infty}{\cal F}_n^{(g)}(a) \epsilon^{2g-2}.}
The first coefficient in this expansion, ${\cal F}_n^{(0)}(a)$,
gives the prepotential of the corresponding ${\cal N}=2$ theory.
It was conjectured in \Nekrasov\ that the remaining coefficients
${\cal F}_n^{(g)}(a)$, $g \ge 1$, give the $n$-instanton
correction to the gravitational couplings ${\cal F}^{(g)}$ of 
the ${\cal N}=2$ theory. Here we provide nontrivial evidence for this
conjecture by comparing some of the instanton computations of
\refs{\Nekrasov,\Flume} with the results of section 2 (and, therefore,
with the matrix model computations).

In \Flume\ ${\cal F}_n (\epsilon_1, \epsilon_2)$ was explicitly
computed for $SU(2)$, up to $n=4$. To compare with our results
in section 2, we restore units as follows: ${\cal F}_n \rightarrow
(\Lambda/2)^{4n} {\cal F}_n$. For $n=1$ one finds
\eqn\fintsone{ \epsilon^{-2}{\cal F}_1 (\epsilon)={\epsilon^{-2}
\over 2^5}{\Lambda^4 \over a^2 }.}
This says that the
one-instanton contribution to ${\cal F}^{(g)}$ vanishes for $g>0$,
which agrees with \swfone\ and \swfhigher. For $n=2$, one finds:
\eqn\finsttwo{ \eqalign{ \epsilon^{-2}{\cal F}_2 (\epsilon)=&
{\Lambda^8 \over 2^8}{ 10 a^2 -  \epsilon^2 \over 8 \epsilon^2 a^4
(4 a^2 - \epsilon^2)^2}\cr =& \epsilon^{-2} {5 \over
2^{14}}{\Lambda^8 \over a^6 } + {1 \over 2^{13}}{\Lambda^8 \over
a^8} + \epsilon^{2} {11 \over 2^{18}} {\Lambda^8 \over a^{10} }  +
\epsilon^{4} {7 \over 2^{19}} {\Lambda^8 \over a^{12} } +
\cdots\cr}}
For $n=3$, one finds from \Flume:
\eqn\fintsthree{ \eqalign{ \epsilon^{-2}{\cal F}_3
(\epsilon)=&{\Lambda^{12} \over 2^{12}} { 18 a^4 -13 a^2
\epsilon^2 +  \epsilon^4 \over 24 \epsilon^2 a^6 (
a^2-\epsilon^2)^2 (4 a^2 - \epsilon^2)^2}\cr =&\epsilon^{-2} {3
\over 2^{18}}{\Lambda^{12} \over a^{10} } + {1 \over 3 \cdot
2^{14}}{\Lambda^{12} \over a^{12}} + \epsilon^{2} {117 \over
2^{22}} {\Lambda^{12} \over a^{14} }  + \epsilon^{4} {293 \over
2^{23}} {\Lambda^{12} \over a^{16} } + \cdots\cr}}
Finally, for $n=4$, one can extract from \Flume:
\eqn\finstfour{\eqalign{\epsilon^{-2}{\cal F}_4(\epsilon)=
&{\Lambda^{16}\over 2^{16}}{23504 a^{10}-70872 a^8\epsilon^2
+67461 a^6\epsilon^4-26339 a^4\epsilon^6+3708 a^2 \epsilon^8
-162\epsilon^{10}\over 128\epsilon^2 a^8
(4 a^2-9\epsilon^2)^2(a^2-\epsilon^2)^2(4 a^2-\epsilon^2)^4}\cr
&=\epsilon^{-2}{1469\over 2^{31}}{\Lambda^{16}\over a^{14}}
+{1647\over 2^{29}}{\Lambda^{16}\over a^{16}}
+\epsilon^2{171201\over 2^{34}}{\Lambda^{16}\over a^{18}}
+\epsilon^4{985823\over 2^{35}}{\Lambda^{16}\over a^{20}}+\dots}}
We see that, after relabelling $\epsilon \rightarrow i
\epsilon$, the coefficients in the above expansions are in perfect
agreement with the results in \swfnull, \swfone\ and \swfhigher.

\vskip1cm

\noindent {\bf Acknowledgements:} 
We thank R.
Flume, S. Gukov, V. Kazakov, I. Kostov, C. Kristjansen and C. Vafa 
for valuable discussions. 
The work of
M.M. was supported by the grant NSF-PHY/98-02709; 
A.K. and S.T. would like to
thank the Erwin-Schr\"odinger-Institut in Vienna, where part of
this work was done, for hospitality. 
They were also partially supported by 
the European Commission TMR programme ERBFMRX-CT96-0045 and 
the work of S.T. in addition by 
GIF, the German-Israeli Foundation for Scientific Research. 

\vfill\eject

\appendix{A}{}

We collect the first few polynomials $g_n^{(g)}$ which enter the topological
amplitudes \growthbehaviour.
\eqn\htwO{
\eqalign{
h^{(2)}_1&=(1-q)^4 \cr
h^{(2)}_2&={1\over 2}(2-17 q + 80 q^2+ 1190 q^3+80 q^4-17q^5+2q^6) \cr
h^{(2)}_3&={1\over 3}(3-38\,q+1712\,q^2+48326\,q^3+124634 q^4+\ldots)\cr
h^{(2)}_4&={1\over 4}(4-67\,q+14592\,q^2+711059\,q^3+4924138\,q^4
+9244668\,q^5+\ldots)\cr
h^{(2)}_5&={1\over 5}(5-104\,q+76705\,q^2+6090098\,q^3
+82568187\,q^4+358139062\,q^5+580752958\,q^6+\ldots)\cr
}}
\eqn\hthree{
\eqalign{
h^{(3)}_1&=(1-q)^6 \cr
h^{(3)}_2&={1\over8}(1+4q+q^2)(8-125q+1016q^2-8854q^3+1016q^4-125 q+8 q^6)\cr
h^{(3)}_3&={1\over 3^3} (3^3-428q+207q^2-584608q^3-6606954q^4-13969512 q^5-\ldots)\cr
h^{(3)}_4&={1\over 4^3} (4^3-1275q-62364q^2-14760656q^3-289184988q^4
-1451906781q^5-2450425504q^6-\dots )\cr
h^{(3)}_5&={1\over 5^3}(5^3-2994q-655802q^2-183946424q^3-5822909304q^4
-53583541710q^5\cr
&\qquad-190175648587q^6-288031344528q^7-\ldots )}}
They were used to derive \swfhigher. Note that the $h^{(g)}_{d_1}$
can be used to write down the conjecturally integer 
Gopakumar-Vafa invariants 
for ${\cal O}(-2,-2)\to\F_0$
(see e.g. \KKVtwo) 
$n^{(g)}_{d_1,d_2}$ for $g<4$, $d_1<6$ and $d_2$ arbitrary, e.g.
$$\eqalign{\{n^{(3)}_{4,d_2},d_2=0,\ldots\}=&
\{0,0,15,4680,184056,3288688,36882969,300668486,\cr
  &1935031484,10359890196,47820549652,195274337280,719145083800,\ldots\}
}$$  
We checked integrality of the $n^{(g)}_{d_1,d_2}$ for $g<4$, $d_1<6$ and $d_2\le 2000$.

\appendix{B}{}

We collect some of the intermediate results which are needed to
derive \Fonematrix. As we mentioned in sect.~4, the derivation is
almost identical to that in sect.~6 of \Akemann. Instead of the
elliptic integrals $K_i$ defined there, we encounter (we choose
$x_1<x_2<x_3<x_4$)
\eqn\Kidef{
K_i=\int_{x_3}^{x_4}{\sqrt{(x-x_1)(x-x_2)(x-x_3)(x-x_4)}\over(x-x_i)}dx}
They arise from varying the constraint that
$N_1=\int_{x_3}^{x_4}\rho(x)dx$ is kept fixed rather than
requiring $\int_{x_2}^{x_3}\rho(x)dx=0$ as in \Akemann.
Explicit calculation gives \BF
\eqn\Kiexplicit{
\eqalign{
K_1&=iX\bigl\lbrace(-\a^6+2\a^4+\a^2 k^2-2\a^4 k^2)E(k)
-(k^2-\a^2)(\a^4-2\a^2+k^2)K(k)\cr
&\qquad\qquad+(\a^8-4\a^6 k^2+6\a^4 k^2-4\a^2 k^2+k^4)
\Pi(\a^2,k^2)\bigr\rbrace\cr
K_2&=iX\bigl\lbrace-\a^2(\a^4-2\a^2-2\a^2 k^2+3 k^2)E(k)
-(k^2-\a^2)(\a^4-2\a^2+4\a^2 k^2-3k^2)K(k)\cr
&\qquad\qquad+(\a^8-6\a^4 k^2+4\a^2 k^4+4\a^2 k^2-3k^4)
\Pi(\a^2,k^2)\bigr\rbrace\cr
K_3&=iX\bigl\lbrace(3\a^6-2\a^4-2\a^4 k^2+\a^2 k^2)E(k)
+(k^2-\a^2)(3\a^4-2\a^2-k^2)K(k)\cr
&\qquad\qquad+(-3\a^8+4\a^6+4\a^6 k^2-6\a^4 k^2+k^4)\Pi(\a^2,k)\bigr\rbrace\cr
K_4&=iX\bigl\lbrace(-\a^6-2\a^4+2\a^4 k^2+\a^2 k^2)E(k)
-(k^2-\a^2)(\a^4+2\a^2-4\a^2 k^2+k^2)K(k)\cr
&\qquad\qquad+(\a^8-4\a^6+6\a^4 k^2-4\a^2 k^4+k^4)\Pi(\a^2,k)\bigr\rbrace}}
where
\eqn\definitions{
\eqalign{
\a^2&={(x_4-x_3)\over(x_4-x_2)}\cr
k^2&={(x_1-x_2)(x_3-x_4)\over(x_1-x_3)(x_2-x_4)}\cr
X&={1\over4}{(x_1-x_3)^{3/2}(x_2-x_4)^{7/2}\over
(x_4-x_3)^2(x_3-x_2)}}}
Next we have to solve eqs. (6.1) in \Akemann\
for the coefficients $\a_i$, however with $K_i$
as given above. We find
\eqn\alphas{
\eqalign{
\a_1&={1\over(x_1-x_2)}\left[1-{(x_4-x_2)\over(x_4-x_1)}{E(k)\over K(k)}
\right]\cr
\a_2&={1\over(x_2-x_1)}\left[1-{(x_3-x_1)\over(x_3-x_2)}{E(k)\over K(k)}
\right]\cr
\a_3&={1\over(x_3-x_4)}\left[1-{(x_4-x_2)\over(x_3-x_2)}{E(k)\over K(k)}
\right]\cr
\a_4&={1\over(x_4-x_3)}\left[1-{(x_3-x_1)\over(x_4-x_1)}{E(k)\over K(k)}
\right]}}
With these values for $k^2$ and $\a_i$, the derivation
of $F^{(1)}$ proceeds exactly as in \Akemann\ and finally leads
to \Fonematrix.

\listrefs

\bye